\documentclass{article}
\usepackage{authblk}
\usepackage[font=small,labelfont=bf,justification=justified]{caption}
\usepackage{float}
\usepackage{mathpazo}
\usepackage{graphicx}
\usepackage[margin=0.7in]{geometry}

\title{\textbf{Doughnut shaped emission from vertical organic nanowire coupled to thin plasmonic film}}

\begin{document}
\author[1]{Adarsh B. Vasista}
\author[1,$\dagger$]{Ravi P. N. Tripathi}
\author[1]{Shailendra K. Chaubey}
\author[1]{Sunny Tiwari}
\author[1,2,$\star$]{G.V. Pavan Kumar}

\affil[1]{\textit{Photonics and Optical Nanoscopy Laboratory, Division of Physics, Indian Institute of Science Education and Research, Pune - 411008, Maharashtra, India}}

\affil[2]{\textit{Center for Energy Science, Indian Institute of Science Education and Research, Pune - 411008, India}}
\affil[$\dagger$]{\textit{Present Address: Jai Prakash Vishwavidyalaya, Chapra - 841301, India}}
\affil[$\star$]{\textit{Corresponding author: pavan@iiserpune.ac.in}}
\maketitle
\date{}


\begin{abstract}
Vertical nanowires facilitate an innovative mechanism to channel the optical field in the orthogonal direction and act as a nanoscale light source. Subwavelength, vertically oriented nanowire platforms, both of plasmonic and semiconducting variety can facilitate interesting far field emission profiles and potentially carry orbital angular momentum states. Motivated by these prospects, in this letter, we show how a hybrid plasmonic - organic platform can be harnessed to engineer far field radiation. The system that we have employed is an organic nanowire made of diaminoanthroquinone grown on a plasmonic gold film. We experimentally and numerically studied angular distribution of surface plasmon polariton mediated emission from a single, vertical organic nanowire by utilising evanescent excitation and Fourier plane microscopy. Photoluminescence and elastic scattering from single nanowire was analysed individually in terms of in plane momentum states of the outcoupled photons. We found that the emission is doughnut shaped in both photoluminescence and elastic scattering regimes. We anticipate that the discussed results can be relevant in designing efficient, polariton-mediated nanoscale photon sources which can carry orbital angular momentum states.             
\end{abstract}




 Engineering the far-field emission has emerged as an important task in realizing miniaturized optoelectronic and nanophotonic platforms \cite{ref1, ref2} especially in the context of single photon sources \cite{ref3}, vertical oriented LEDs \cite{ref4,ref35} and other optical communication devices \cite{ref2, ref2_1}. Effective manipulation of out-coupled optical field and its directionality can pave the way in improving the existing optoelectronic/photonic devices \cite{ref2_1, ref5, ref6}. Several efforts are reported to control the emission at subwavelength scale by designing various plasmonic \cite{ref7, ref8, ref9, ref10,ref39,ref38}, alternative plasmonic materials \cite{ref11}, molecular nanoarchitectures \cite{ref6,ref12,ref13,ref14,ref15} etc. Furthermore, several hybrid schemes have been demonstrated to tailor the radiative properties of nanostructures \cite{ref16,ref17,ref18,ref19}. Among many, plasmon - organic hybrid scheme  has gained significant attention in recent years and is employed to realize various hybrid features \cite{ref18,ref19,ref20}. \\

	Recently, we have shown that a vertically oriented, single organic nanowire coupled to a plasmonic thin film can be used as sub-wavelength exciton-polariton source \cite{ref19} whose emission intensity can be tuned by plasmon - exciton coupling efficiency. Furthermore, we have shown that similar system can radiatively channel exciton-polaritons through surface plasmons \cite {ref20}. An important aspect that is yet to be explored in such a system is the far-field intensity pattern of the out-coupled radiation. Given that vertically oriented nanowire structures can result in interesting radiation patterns \cite {ref4,ref12,ref12_1,ref41,ref44}, polarization \cite{ref40,ref30} and near field profiles \cite{ref42}, studying them will have direct implication on sub-wavelength light emitting devices \cite{ref32,ref33}, resonators \cite{ref34} and nano-optical sources carrying orbital angular momentum \cite{ref31,ref46,ref47,ref45}.\\

\begin{figure}[H]
\centering
\includegraphics{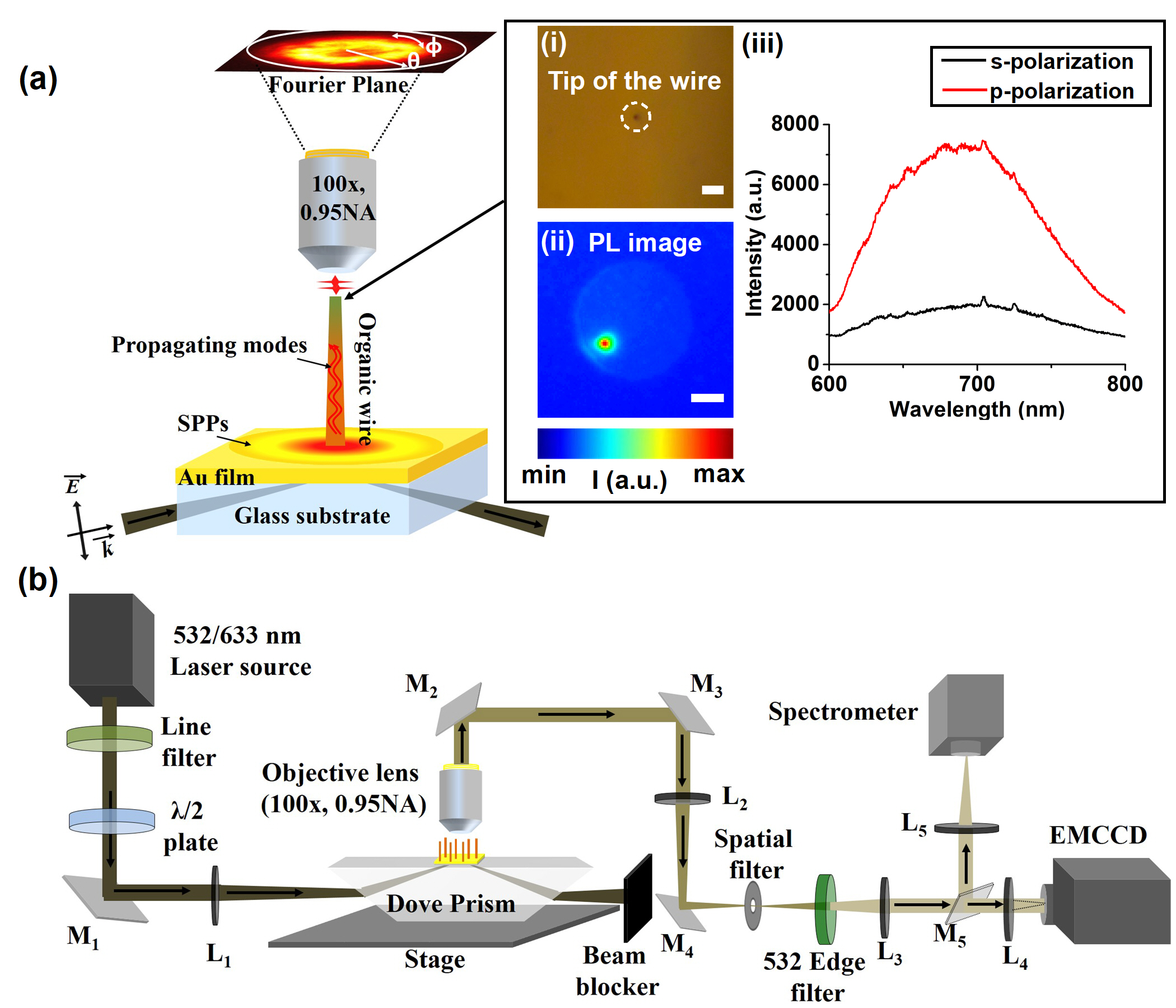}
\caption{ (a) Conceptual schematic of plasmonic (Au) film coupled single, vertical organic (DAAQ) nanowire. Surface plasmon polaritons (SPPs) in Au-film (thickness$\sim$50nm) were generated using evanescent exitation with 532 nm / 633 nm laser beam. This field facilitates the excitation of the propagating modes in the organic nanowire, which outcouple at the distal end of the wire. By changing the excitation wavelength from 532 nm to 633 nm, we switch between active and passive operating regimes of the nanowire. The light from the distal end was collected and projected into the Fourier plane and analyzed. Figure not to scale. \textbf{Insets} : (i) Real plane brightfield image of the tip of the organic nanowire. (ii) Real plane photoluminescence image captured by focusing the tip of the nanowire. (iii) Photoluminescence spectra collected from an individual organic nanowire for different input polarizations. Scale bars are 2$\mu$m. (b) Schematic representation of optical microscopy setup for performing both conventional and Fourier plane imaging.}
\label{fig:1}
\end{figure}
	
	Motivated by this, we studied experimentally and numerically the resulting radiation pattern from the distal, free end of a single, vertically oriented organic nanowire made of 1,5-Diaminoanthroquinone (DAAQ) coupled to Au thin film. DAAQ nanowires can be operated in two regimes: Active regime, where the emission is dominated by the Frenkel exciton-polaritons \cite{ref36} and passive regime, where the wire behaves as a dielectric waveguide. These regimes can be probed by using suitable excitation wavelengths. Active emission regime, leading to photoluminescence, can be probed by exciting the DAAQ wire at the absorption maximum, $\sim$ 500 nm at the excitation wavelength, of the wire \cite{ref15}. The input photons will create Frenkel exciton-polaritons in the wire and outcouple at the distal end as photoluminescence. Thus the output light will be red shifted as compared to the input wavelength. If we excite the wire away from the absorption maximum, the wire acts as a dielectric waveguide and can be used to study light propagation and emission at sub wavelength limit. Here the output light from the distal end of the nanowire will be of the same wavelength as that of the input light. This operational regime tunability of organic nanowire makes it different from its plasmonic counterparts\cite{ref37,ref38,ref30}. \\

	 Harnessing this flexibility of the DAAQ nanowire, we study the radiation pattern of such organic nanowires grown vertically over a metallic film both in active (photoluminescence) and passive (elastic scattering) operational regimes. In this report, we show that the far-field emission pattern of an individual nanowire is doughnut shaped. All these results indicate that organic nanowire on a plasmonic film can be potentially harnessed as a platform to generate emission in the form of vectorial vortex beams. \\
	
	Figure \ref{fig:1}(a) shows the schematic of the experiment. Hybrid nanostructures were prepared via physical vapour transport method. Detailed protocol to prepare these structures can be found in our previous reports \cite{ref19,ref20}. Surface plasmons in the film can be efficiently excited using evanescent excitation, which inturn couples the energy to the vertical wire by exciting propagating modes.  The coupled light travel through the length of the wire and outcouples radiatively. Insets (i) and (ii) of figure \ref{fig:1} show real plane brightfield and photoluminescence images captured by focusing the tip of an individual organic nanowire. Photoluminescence spectral information collected from the vertical nanowire shows very strong input polarization dependence as shown in inset (iii) of figure \ref{fig:1}(a). When the input was \textit{p}-polarized, we get considerably more photoluminescence compared to \textit{s}-polarized input. Passive emission also shows strong elastic scattering for \textit{p}-polarization (data not shown here). This was mainly due to the fact that the efficiency of excitation of surface plasmons on the film is large for \textit{p}-polarized light. This inturn couples to the propagating mode of the wire, hence showing pronounced emission. This makes the vertical nanowires act as input polarization sensitive switch. \\ 
	
	 Given these facts, an important question in this context is what is the wavevector distribution of outcoupled radiation in both active and passive regime, for plasmon coupled organic hybrid system? To answer this, we performed Fourier plane (FP) microscopy on individual vertical nanowire coupled to Au thin film. The excitation wavelength was chosen corresponding to the operational regimes: 532 nm for active regime (exciting near absorption maximum, as the absorption maximum of DAAQ wires is around 520 nm \cite{ref15}), to probe photoluminescence and 633 nm for passive regime (exciting away from absorption maximum \cite{ref15}), to probe elastic scattering. The light outcoupled from a single vertical wire was collected using an high numerical aperture objective lens and projected into the Fourier plane.\\

	  We experimentally probed emission from vertical organic wire on gold film using home built Fourier microscope. Figure \ref{fig:1} (b) shows the schematic of optical setup. To ensure the single wavelength excitation, we introduced laser line filter in the excitation path (for both 532 nm and 633 nm excitation). Incident beam polarization was varied by using $\lambda/2$ plate in the excitation path. The outcoupled light from a single vertical organic nanowire was captured using high NA (100x, 0.95) objective lens using a spatial filter placed in the conjugate image plane and routed to engaged EMCCD/spectrometer by utilizing 4-$f$ configuration. 4-$f$ transformation was performed using the lens combination of $L_2$ and $L_3$. Real plane imaging was performed by introducing flippable lens $L_4$. The spectral signature of emission was recorded by routing the captured light towards spectrometer using flippable mirror $M_5$. For the photoluminescence microscopy, elastically scattered light was rejected by introducing 532nm edge filter. With 633 nm laser excitation, the outcoupled light is dominated by elastic scattering with practically no contribution from photoluminescence \\
	   
	  To understand the emission process further , we performed finite element method (FEM) based full-wave numerical simulation using COMSOL software (version 5.1). Near field to far field transformation were carried out using reciprocity arguments \cite{ref26}. Nanowire geometry was taken as a tapered cone with dimensions derived from SEM images, mimicking the realistic situation. DAAQ wires were modelled using lorentzian oscillator model and wavelength dependent refractive indices were taken from \cite{ref15, ref49} and that of gold from \cite{ref27}. The radius of the base of the wire was taken to be of 300 nm and that of the tip to be 100 nm, thus making the wire to have a continuous taper. The wire was placed on a 50 nm gold film over the glass substrate. To simulate the photoluminescence emission from vertical nanowires, vertical dipole source oscillating at 700 nm (wavelength at which photoluminescence emission is maximum)was placed inside the nanowire at nanowire - Au film interface \cite{ref19,ref12}. The simulation region was terminated by scattering boundary conditions to avoid spurious reflections. The system was meshed using free tetrahedral mesh with maximum element size of 20 nm to ensure accuracy. The light out coupled from the nanowire was collected and projected to the far field. To model elastic scattering from vertical nanowires, the vertical dipole oscillating at 633 nm was placed at the gold - glass interface right below the nanowire \cite{ref48}. The oscillating dipole excites the propagating plasmons in the film, which inturn excites the wire. The light outcoupled from the wire was collected and projected to the far field.\\

\begin{figure}[H]
\centering
\includegraphics{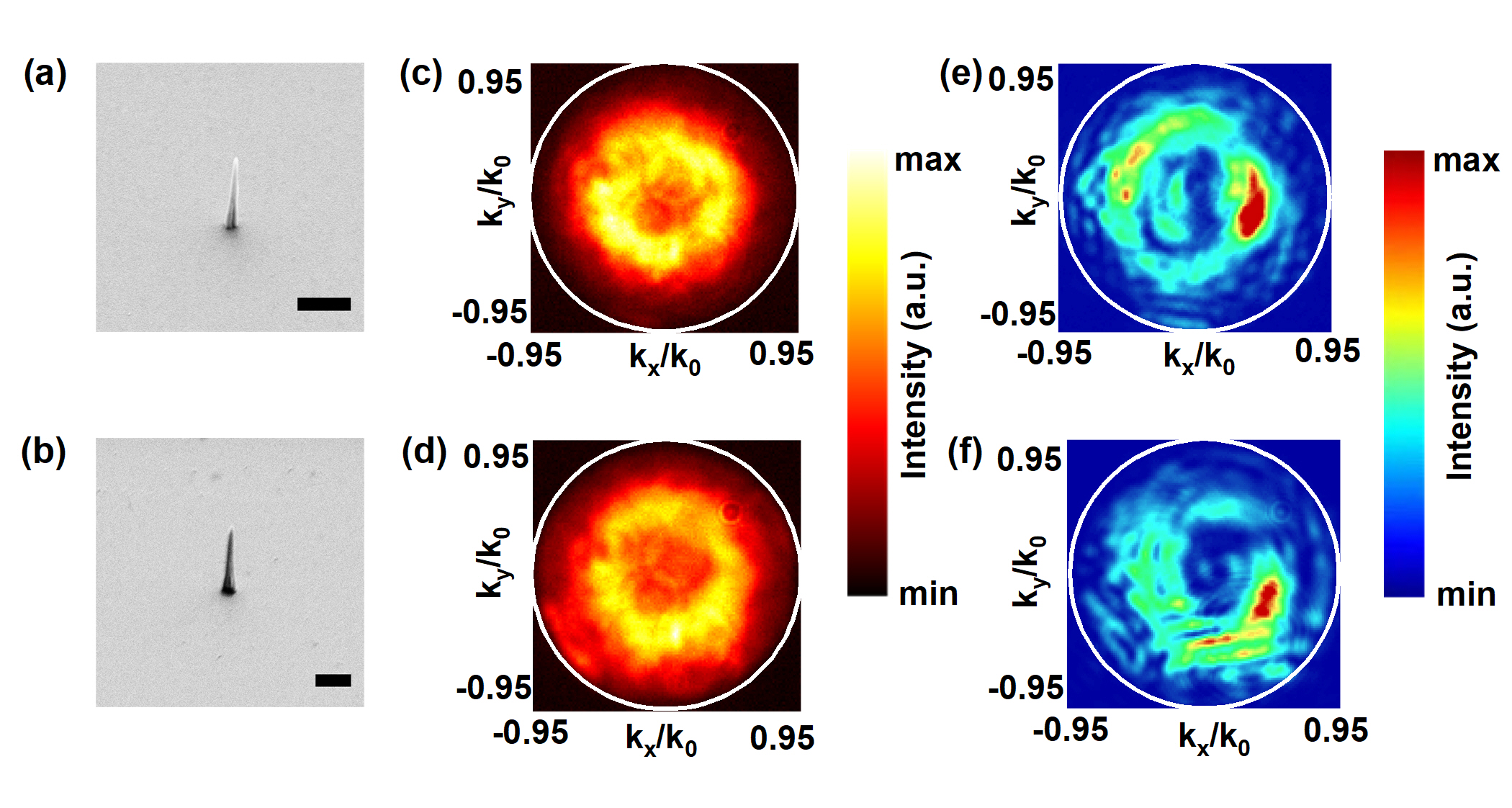}
\caption{(a) and (b) are representative scanning electron micrographs showing vertical nature of the organic nanowires with respect to the substrate. Scale bar is 2 $\mu$m. The wires are typically 2-3$\mu$m in length and the radius at the tip is $\sim$ 200 nm. (c) and (d) are photoluminescence Fourier plane images captured from two individual nanowires shwoing doughnut emission pattern. (e) and (f) are corresponding elastic scattering Fourier plane images.  }
\label{fig:2}
\end{figure}
	 
	Figure \ref{fig:2}(a) and (b) show the representative SEM images of the vertical organic nanowires on Au film. They clearly indicate that the wires are grown vertically over the substrate. The typical length of the wire was $\sim$ 2-3 $\mu$m. The base radius of the wire was $\sim$ 300 nm and that of the tip was $\sim$ 100 nm, thus making the wire a continuous taper. Equipped with this, we collected the outcoupled light from an individual organic wire and projected it into the Fourier plane. The vertical nanowires on thin Au film were excited using \textit{p}- polarized laser beam through evanescent excitation. We performed measurements with 532 nm and 633 nm laser separately on the same nanowire to probe photoluminescence and elastic scattering FP images. Figures \ref{fig:2} (a) and (b) show the measured photoluminescence FP images from two individual nanowires. We can see clear doughnut like pattern in both the images. This emission pattern is repeatable, as long as wires are vertical to the substrate and the images shown here are representative in nature. Further we measured elastic scattering FP images by same wires by exciting them using 633 nm laser beam. Figures \ref{fig:2}(c) and (d) show the corresponding FP images. It is interesting to note that the elastic scattering FP images are qualitatively similar to the corresponding photoluminescence FP images. Due to spurious scattering from the substrate, the elastic scattering FP image has irregular speckles. But the overall feature of doughnut is clearly visible. \\ 
	
	Doughnut type emissions are seen typically in dipole over mirror configurations\cite{ref28,ref29,ref43}, which is further affected by the presence of an organic semiconductor medium, here. The nanowire acts as a waveguide for the propagating modes of the organic wire. The free end of the nanowire now acts as an effective radiating dipole orienated along the \textit{z}-axis and outcouples the guided modes. \\ 	
	
	\begin{figure}[H]
\centering
\includegraphics{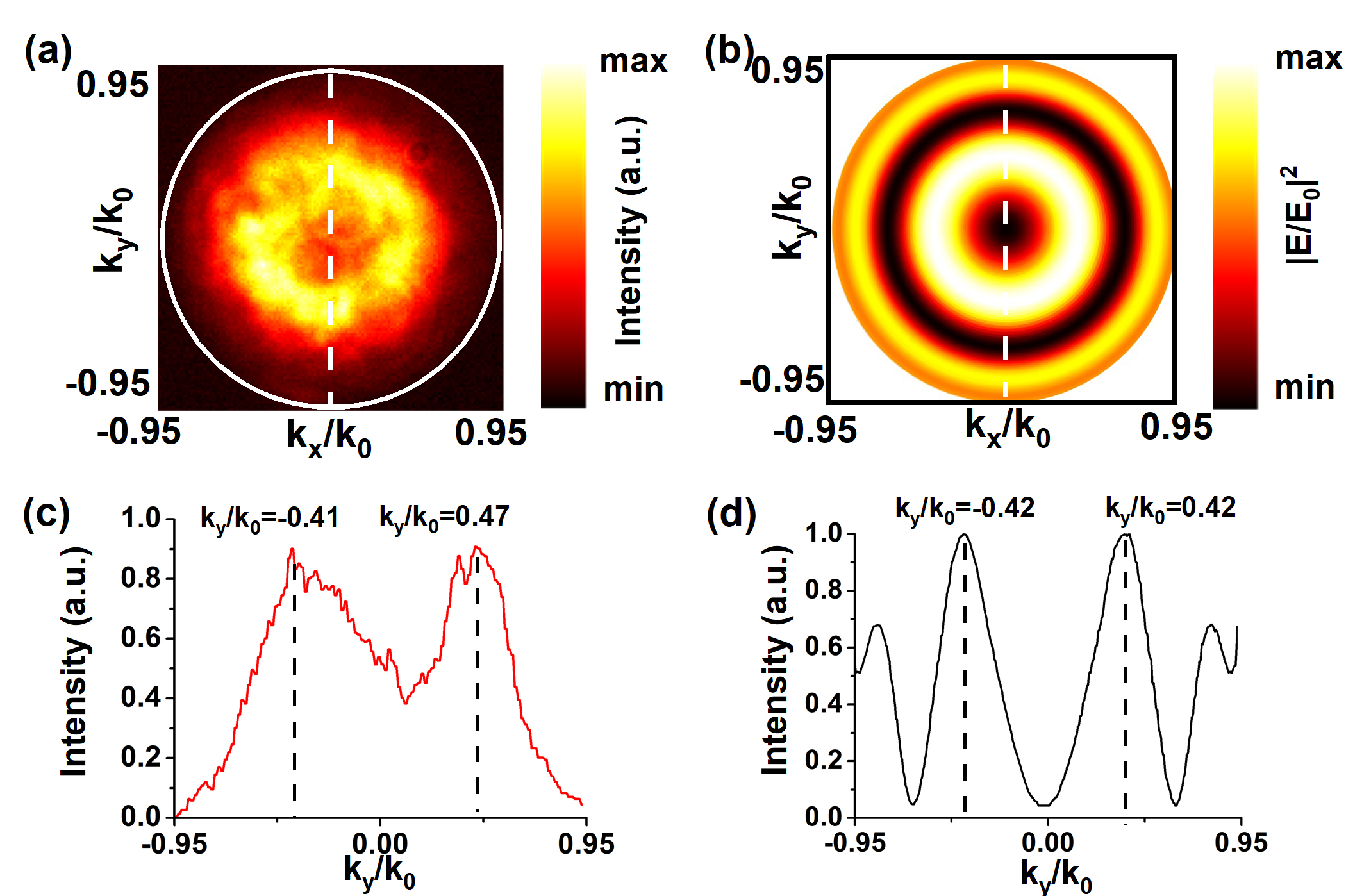}
\caption{(a) Measured Photoluminescence Fourier plane image captured from an individual vertical organic nanowire, showing doughnut typed emission. (b) Simulated photoluminescence Fourier plane image captured from individual vertical organic nanowire at 700 nm wavelength. (c) and (d) are intensity profiles along the dashed line shown in Fourier plane images.}
\label{fig:3}
\end{figure}	
		 	
	Figures \ref{fig:3} (a) and (b) show the measured and simulated photoluminescence FP images respectively. The simulated FP image is similar to the measured FP pattern. To further quantify the emission, we further plot the intensity profiles of the emission along the dashed line for \textit{k$_x$}=0 (figure \ref{fig:3} (a) and (b)) and are shown in figure \ref{fig:3} (c) and (d) respectively. The emission is maximum near \textit{k$_y$/k$_0$}=-0.41 and 0.47 for measured photoluminescence emission. Also emission occurs at \textit{k$_x$/k$_0$}=-0.44 and 0.42 (graph not shown). For the simulated FP image, emission occurs at \textit{k/k$_0$}= $\pm$0.42, which is in good agreement with the measured photoluminescence angular emission values. \\   
	
\begin{figure}[H]
\centering
\includegraphics{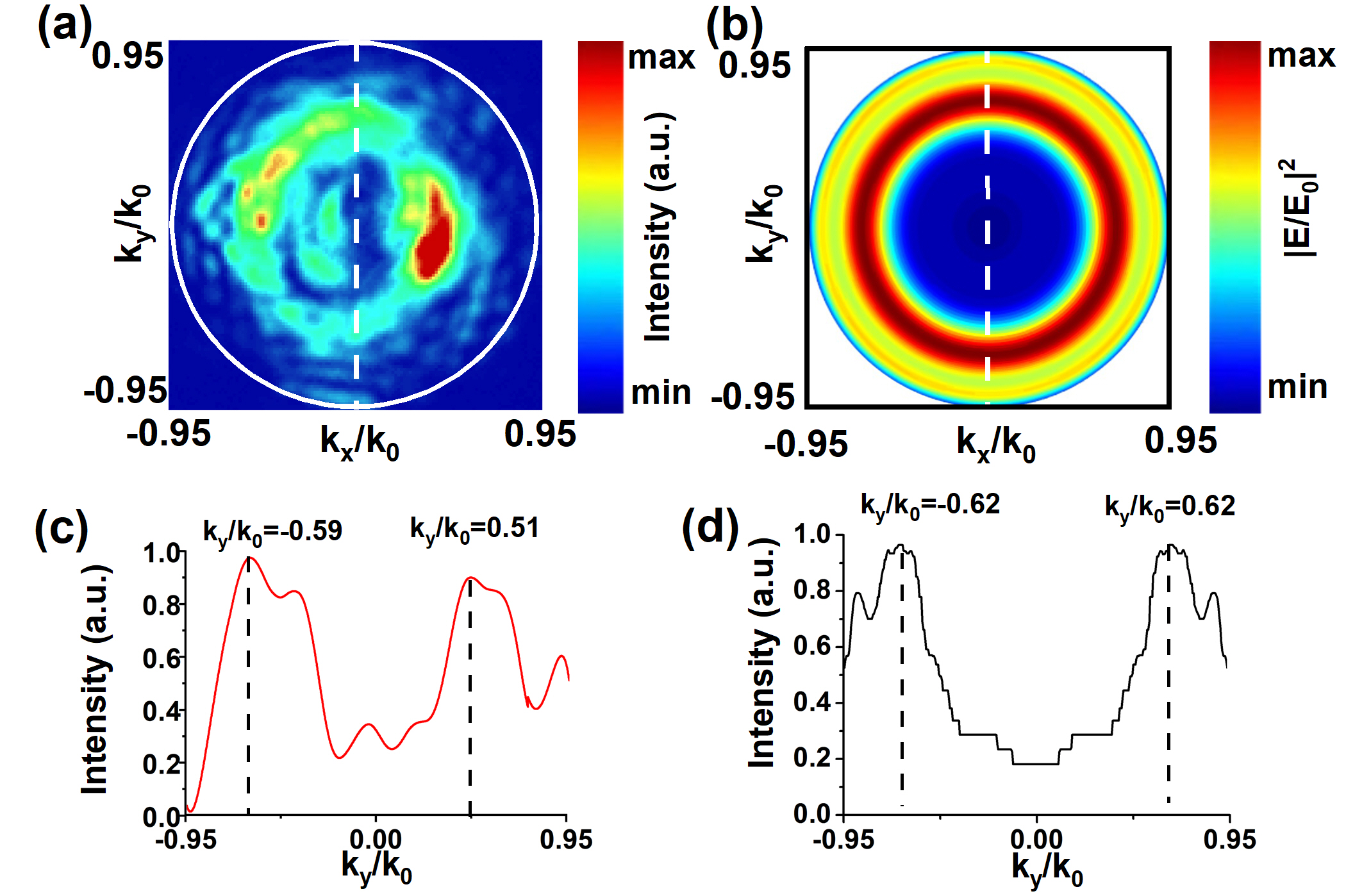}
\caption{(a) Measured elastic scattering Fourier plane image captured from an individual vertical organic nanowire, showing doughnut typed emission. (b) Simulated elastic scattering Fourier plane image captured from individual vertical organic nanowire at 633 nm wavelength. (c) and (d) are intensity profiles along the dashed line shown in Fourier plane images.  }
\label{fig:4}
\end{figure} 

	Figures \ref{fig:4} (a) and (b) show measured and simulated elastic scattering FP images respectively. The angle of scattering were calculated by plotting the intensity profiles along the dashed line in figures \ref{fig:4} (a) and (b). Experimentally measured FP image shows that the  \textit{k}-vectors of scattering are \textit{k$_y$/k$_0$}= -0.59 and 0.51. Also emission occurs at \textit{k$_x$/k$_0$}=-0.54 and 0.46 (graph not shown). For simulation FP images, maximum elastic scattering scattering happens at \textit{k/k$_0$} = $\pm$0.62 which is similar to the values observed in the measurements. So qualitatively photoluminescence and elastic scattering FP images show similar FP images with slightly different emission angles, which warrants further investigation. The FP images show a slight asymmetry which may be due a small tilt in the vertical nanowire with respect to the normal to the substrate.\\ 
	
	By measuring the wavevector distribution of multiple vertical nanowires, we calculated the average emission vectors \textit{k$_x$/k$_0$} and \textit{k$_y$/k$_0$} for both photoluminescence and elastic scattering. For photoluminescence average value of emission wavevector was \textit{k$_x$/k$_0$}=-0.425$\pm$0.02 ; 0.435$\pm$0.01 and \textit{k$_y$/k$_0$}=-0.423$\pm$0.03 ; 0.435 $\pm$0.01. For elastic scattering it was \textit{k$_x$/k$_0$}=-0.482$\pm$0.06 ; 0.5$\pm$0.04 and \textit{k$_y$/k$_0$}=-0.53$\pm$0.08 ; 0.56$\pm$0.02. The values are similar to that observed in the numerical simulations. \\

	To conclude, we experimentally and numerically demonstrate doughnut shaped emission from a single vertical organic nanowire excited via surface plasmons. We studied Au film coupled vertical organic nanowires in both active and passive operational regimes and corroborated our experimental observations with numerical simulations. We anticipate that the obtained results can have implications in realizing efficient light sources carrying orbital angular momentum. These organic wires can be driven to get doughnut pattern in both coherent scattering and incoherent emission regimes. Hence it provides unique opportunity to devise nano optical sources which can carry angular momentum states in two different operational regimes. Simultaneously, the same system can be further extrapolated to probe the quantum (weak or strong) spin-orbit coupling effects in plasmon - organic hybrid system. 
\section*{Acknowledgement}
This research work was partially funded by DST-Nanomission Grant (SR/NM/NS-1141/2012(G)), Center For Energy Science (SR/NM/TP-13/2016). Authors thank Dr. Arindam Dasgupta, Dr. Rohit Chikkaraddy, Vandana Sharma and Deepak K. Sharma for fruitful discussions. ABV and ST thank Infosys foundation, India for the financial aid.

\bibliographystyle{plain}
\bibliography{ref}
\end{document}